\documentclass[letterpaper, 10 pt, conference]{ieeeconf}  

\IEEEoverridecommandlockouts                              

\usepackage{graphics} 
\usepackage{epsfig} 
\usepackage{mathptmx} 
\usepackage{times} 
\usepackage{newtxtext} 

\usepackage{amsmath, amssymb, graphicx}
\usepackage{amsfonts}
\usepackage{algorithmic}
\usepackage{algorithm}
\usepackage{array}
\usepackage{xcolor}
\usepackage{textcomp}
\usepackage{stfloats}
\usepackage{url}
\usepackage{verbatim}
\usepackage{graphicx}
\usepackage{textcomp}
\usepackage{multirow}
\usepackage{siunitx}
\usepackage{comment}
\usepackage{bm}
\usepackage{float}
\usepackage{wrapfig}
\usepackage{cleveref}
\usepackage{booktabs}

\title{\LARGE \bf \emph{SSL-SE-EEG}: A Framework for Robust Learning from Unlabeled EEG Data with Self-Supervised Learning and Squeeze-Excitation Networks}

\author{Meghna Roy Chowdhury, Yi Ding, and Shreyas Sen  
\thanks{Meghna Roy Chowdhury, Yi Ding, and Shreyas Sen are with Elmore Family School of Electrical and Computer Engineering, Purdue University, USA
        {\tt\small \{mroycho,yiding,shreyas\}@purdue.edu}}%
}

\begin{document}

\maketitle
\thispagestyle{empty}
\pagestyle{empty}


\begin{abstract}
Electroencephalography (EEG) plays a crucial role in brain-computer interfaces (BCIs) and neurological diagnostics, but its real-world deployment faces challenges due to noise artifacts, missing data, and high annotation costs. We introduce \emph{SSL-SE-EEG}, a framework that integrates Self-Supervised Learning (SSL) with Squeeze-and-Excitation Networks (SE-Nets) to enhance feature extraction, improve noise robustness, and reduce reliance on labeled data. Unlike conventional EEG processing techniques, \emph{SSL-SE-EEG} transforms EEG signals into structured 2D image representations, suitable for deep learning. Experimental validation on MindBigData, TUH-AB, SEED-IV and BCI-IV datasets demonstrates state-of-the-art accuracy (91\% in MindBigData, 85\% in TUH-AB), making it well-suited for real-time BCI applications. By enabling low-power, scalable EEG processing, \emph{SSL-SE-EEG} presents a promising solution for biomedical signal analysis, neural engineering, and next-generation BCIs. 

\end{abstract}

\section{Introduction}

Electroencephalography (EEG) is a vital biopotential signal used to measure brain activity in applications such as brain-computer interfaces, cognitive monitoring, and the diagnosis of neurological disorders~\cite{hosseini2020review}. Despite its importance, real-world EEG applications face significant challenges due to noise, motion artifacts, and incomplete data from missing or corrupted channels, often resulting from electrode displacement~\cite{puce2017review,gondran1996noise}.

\begin{figure}[ht]
    \centering
    \includegraphics[width=1\linewidth]{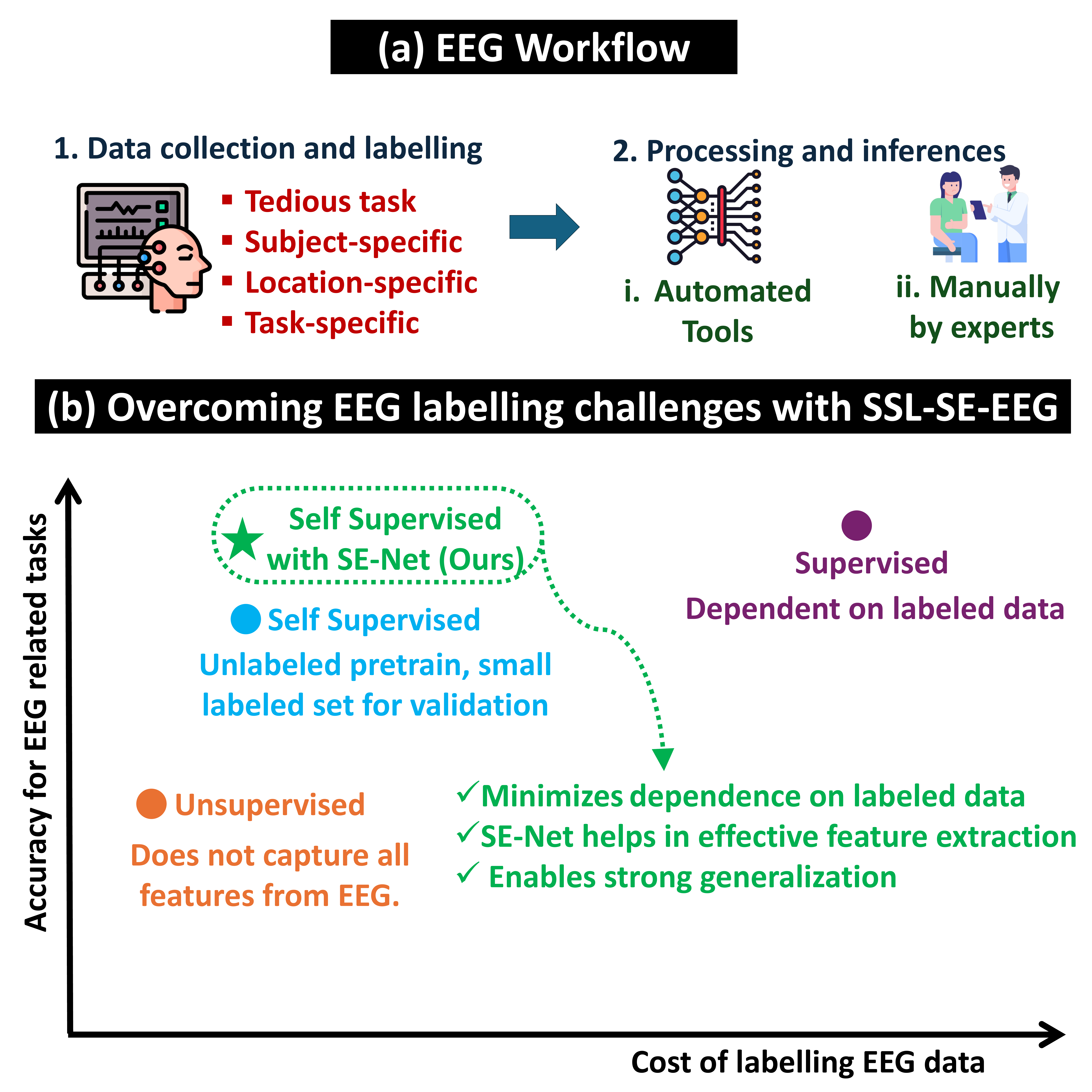}
    \vspace{-8mm}
    \caption{(a) EEG workflow highlighting challenges in data labeling and processing. (b) Trade-off between model accuracy and labeling cost across ML paradigms, showing how \emph{SSL-SE-EEG} achieves high performance with reduced labeled data.}
        \vspace{-5mm}
    \label{fig_intro}
\end{figure}

The traditional  EEG workflow is illustrated in Fig.~\ref{fig_intro}(a). It begins with data collection and labor-intensive labeling, which is subject-specific, task-specific, and highly variable across acquisition conditions. Subsequently, the collected data is processed either manually by domain experts or through automated methods. Automated approaches typically leverage machine learning (ML) techniques, which can be broadly categorized into two groups. Traditional ML techniques include artificial neural networks (ANN), support vector machines (SVM), and principal component analysis (PCA)~\cite{guerrero2021eeg}. More recently, deep learning architectures, particularly convolutional neural networks (CNNs), have shown superior performance, largely due to their ability to effectively extract features from EEG signals~\cite{saeidi2021neural}.

While traditional ML techniques and most CNN-based methods have demonstrated success in EEG processing, they predominantly rely on the supervised learning (SL) paradigm, which requires extensive labeled datasets to uncover meaningful patterns~\cite{nafea2022supervised}. This dependence on manual annotation renders data preparation expensive, time-consuming, and subject to strict human research constraints~\cite{team2020deep,jaiswal2020survey,liu2021emotion}. Moreover, even with ample labeled data, achieving robust generalization across subjects and sessions remains a persistent and unresolved challenge~\cite{kohan2020interview}. Unsupervised learning, which operates solely on unlabeled data, alleviates the burden of annotation but often struggles to learn features that reliably separate signal from noise~\cite{hosseini2020review}. In response to these limitations, self-supervised learning (SSL) has emerged as a promising middle ground. SSL frameworks leverage large volumes of unlabeled data to learn rich, transferable representations through carefully designed pretext tasks, requiring only a small fraction of labeled data for downstream fine-tuning~\cite{gui2024survey}. As shown in Fig.~\ref{fig_intro}(b), SSL offers an attractive trade-off, achieving high performance with substantially reduced labeling costs.

In this paper, we introduce \emph{SSL-SE-EEG}, a novel framework that combines SSL with CNNs and Squeeze-and-Excitation Networks (SE-Nets) to enhance feature extraction and improve robustness to noise in EEG-based image representations. Our approach first transforms EEG signals into structured 2D RGB image representations, preserving critical temporal and amplitude information, thereby enabling compatibility with CNN-based encoders and facilitating diverse view generation for contrastive learning. SE-Nets further refine feature learning by dynamically recalibrating channel-wise responses, allowing the model to focus on the most informative aspects of EEG signals. As highlighted in Fig.~\ref{fig_intro}(b), \emph{SSL-SE-EEG} delivers high classification accuracy while significantly minimizing reliance on labeled data. 

We validate our framework across four public EEG datasets, demonstrating strong generalization across subjects and tasks. Additionally, we show that SE integration introduces minimal power overhead, making \emph{SSL-SE-EEG} particularly well-suited for deployment in low-power, wearable EEG systems~\cite{chowdhury2024leveraging,chatterjee2023bioelectronic,sen2024human}.

We summarize our main contributions as follows:
\begin{itemize}
    \item \textbf{\emph{SSL-SE-EEG} Framework:} We propose a novel framework that integrates self-supervised learning with Squeeze-and-Excitation Networks to enhance feature extraction, improve generalization, and reduce dependency on labeled EEG data.
    
    \item \textbf{EEG Representation as 2D Images:} We transform EEG signals into 2D RGB image representations that preserve temporal and amplitude characteristics, enabling effective feature learning through CNN.
    
    \item \textbf{Accuracy and Energy Efficiency with SE-Nets:} We validate \emph{SSL-SE-EEG} on multiple public EEG datasets, achieving high classification accuracy while maintaining lightweight SE-Net integration, ensuring suitability for energy-efficient, wearable EEG applications.
\end{itemize}

\section{Relevant Topics to Understand}
\begin{figure*} [t]
    \centering
    \includegraphics[width=0.8\linewidth]{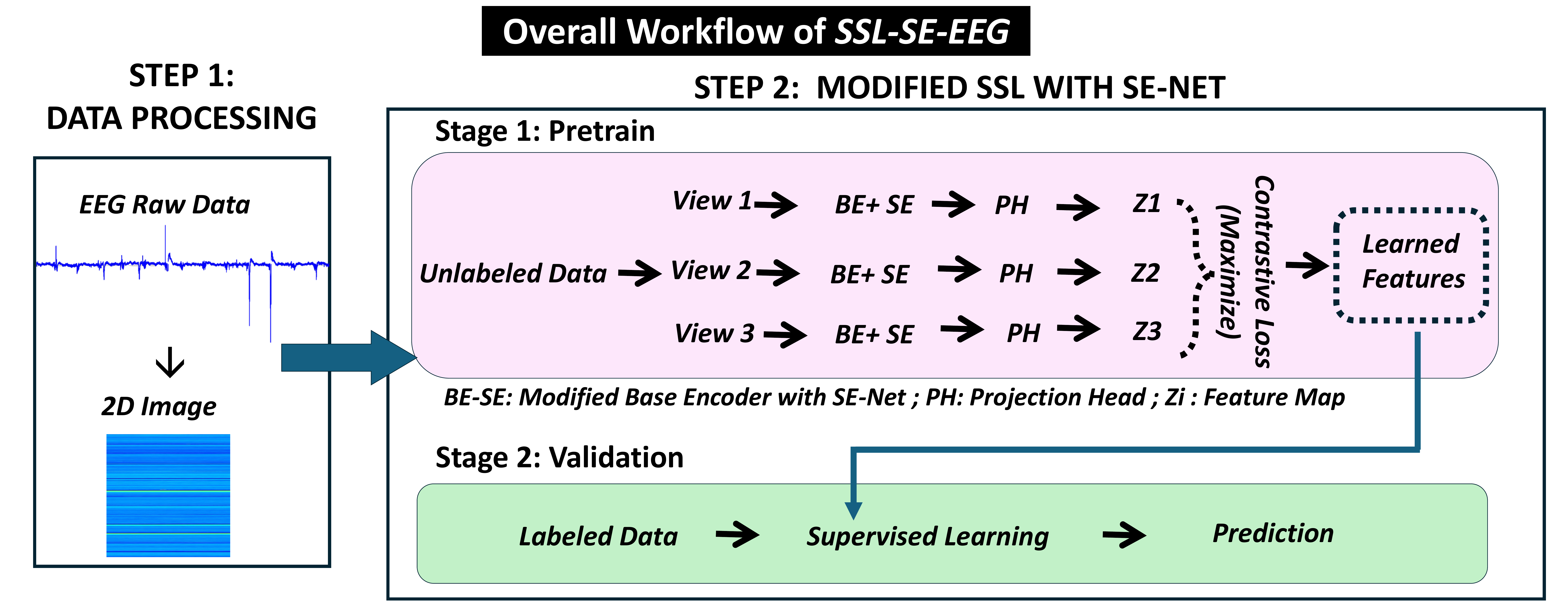}
    ~\vspace{-3mm}
    \caption{Overview of the proposed \emph{SSL-SE-EEG} pipeline, which consists of two steps. Step 1 involves preprocessing EEG signals into 2D image representations. Step 2 integrates self-supervised learning with SE-Nets through a two-stage process: Stage 1 applies contrastive learning using a modified encoder (BE+SE) and a projection head (PH) to maximize feature diversity; Stage 2 fine-tunes the network on a smaller labeled dataset for classification, leveraging learned representations for robust inference.}
    ~\vspace{-6mm}
    \label{fig_workflow}
\end{figure*}

\subsection{Self-Supervised Learning (SSL)}

SSL is a machine learning paradigm in which models are trained on unlabeled data using inherent structures or generating supervisory signals through pretext tasks~\cite{gui2024survey}. In summary, SSL methods typically design auxiliary tasks, such as predicting missing segments, contrasting different views of the same data, or reconstructing inputs, to force the model to learn valuable representations without relying on manual annotations~\cite{jaiswal2020survey}. Recent advances include techniques such as contrastive learning (e.g. SimCLR~\cite{chen2020simple}, MoCo~\cite{he2020momentum}) and masked autoencoders, which have significantly improved performance in areas such as computer vision\cite{ohri2021review,ramesh2023dissecting,hojjati2024self} and natural language processing~\cite{elnaggar2021prottrans,baevski2023efficient,morais2022speech}. SSL has been used in biomedical datasets like knee MRI, SARS-COV-CT, and TissueMNIST and has shown promising results~\cite{atito2022sb,tan2024self,huang2024systematic}. However, studies on EEG and SSL are limited~\cite{mohsenvand2020contrastive,banville2021uncovering}.

 \subsection{Squeeze and Excitation Network (SE-Net)}

SE-Net is a neural network architecture designed to improve feature representations by explicitly modeling the interdependencies between channels~\cite{hu2018squeeze}. It works by "squeezing" global spatial information into a channel descriptor through a global pooling layer and then "exciting" or recalibrating these channels via a gating mechanism with fully connected layers, allowing the network to focus on the most informative features. Recent work has integrated SE-Net modules into various architectures to improve performance in tasks such as image classification, object detection, and semantic segmentation~\cite{sun2021squeeze,xu2022scale,wang2022superpixel}. Moreover, its adaptability has shown promise in the biomedical domains, where channel-wise recalibration can help to capture better and emphasize critical information~\cite{ge2021convolutional,xiong2024sea,li2021msgse,hayat2024squeeze}.

Despite their success in other domains, SE-Nets have been limited in use in EEG analysis. This work integrates SSL and SE-Nets to create a label-independent framework that adapts to missing channels and efficiently learns the appropriate features of EEG. The following section details our methodology, outlining how SSL and SE-Nets improve feature learning and generalization in EEG processing.

\begin{figure*}[t]
\centering
    \includegraphics[width=1\linewidth]{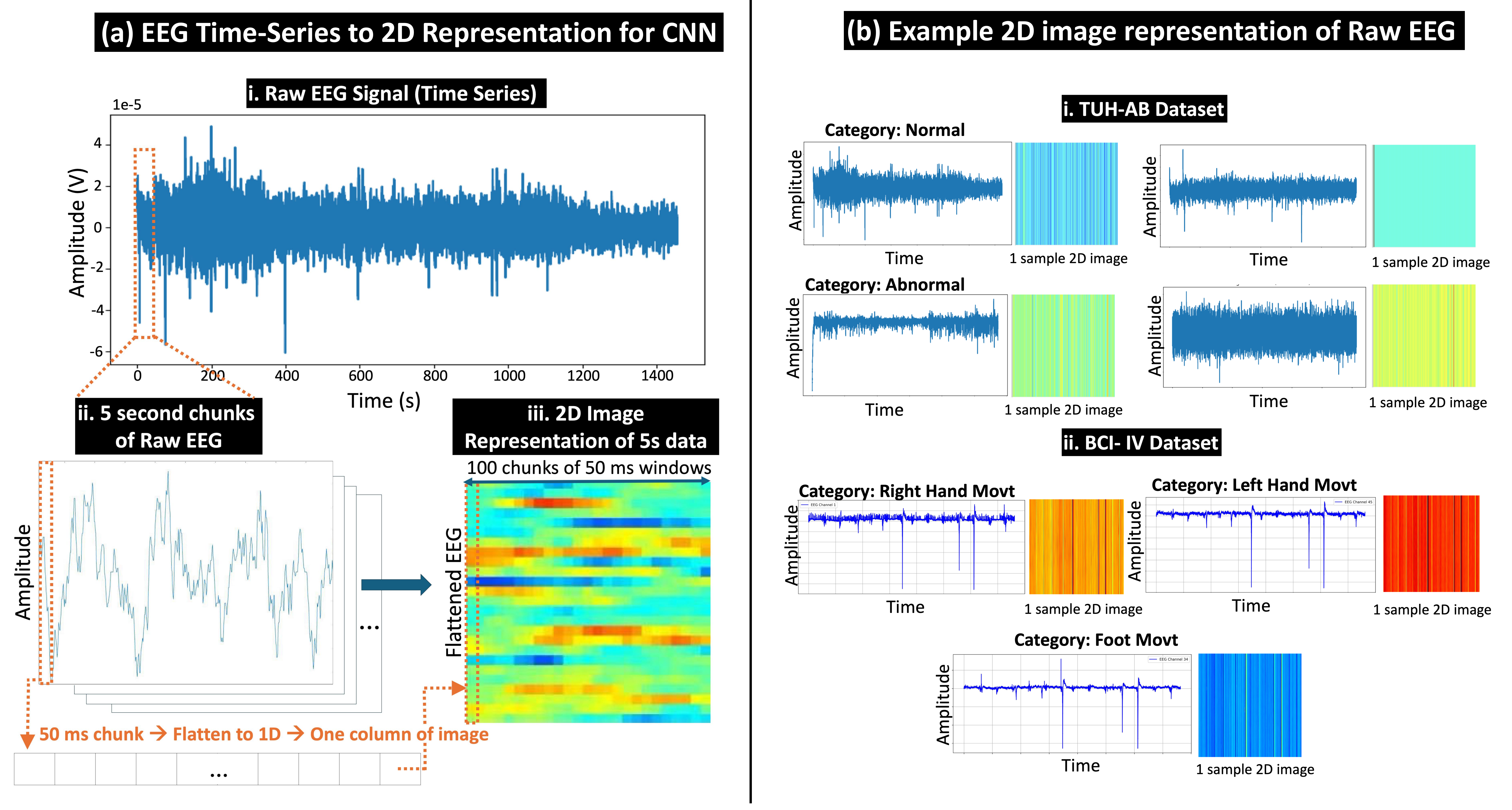}
    ~\vspace{-5mm}
    \caption{Illustration of EEG-to-2D image conversion: (a) Raw EEG signals are segmented into 2-second windows, where 50 ms temporal segments are flattened into 1D columns to construct a 2D image. (b) Examples from TUH-AB and BCI-IV datasets demonstrate the effectiveness of this transformation for CNN-based feature extraction and classification.}
    ~\vspace{-4mm}
    \label{fig_dataprep}
\end{figure*} 

\section{Proposed framework - \emph{SSL-SE-EEG}}
\label{sec_method}

To reduce the reliance on labeled data while enhancing robustness and generalization, we introduce Self-Supervised Learning with Squeeze-and-Excitation Networks for EEG \emph{SSL-SE-EEG}. This framework enables robust feature extraction from unlabeled EEG signals while dynamically enhancing relevant patterns and suppressing noise. Additionally, it leverages a CNN-friendly 2D image representation of EEG data, making it well-suited for deep learning architectures. Fig.~\ref{fig_workflow} outlines the workflow. It starts with \emph{Data Processing}, followed by the \emph{the modified SSL implementation with SE-Nets}. Each step is described below.

\subsubsection{\textbf{Data Processing}} 
Raw EEG signals are time-series waveforms that may not be directly compatible with many deep learning architectures predominantly designed for image-based inputs. To bridge this gap, we introduce a procedure to transform EEG waveforms into a structured 2D image representation, preserving temporal and amplitude information in a format that CNNs can effectively process. The concept of transforming raw signal data into image representations for CNN processing has been previously explored in different contexts such as RF signal analysis~\cite{bari2023rf}, and finds use here for EEG signals. The process is illustrated in Fig.~\ref{fig_dataprep}(a).

We begin by segmenting the raw EEG into 5-second windows, ensuring that each generated image encapsulates a fixed duration of EEG activity. Each 5-second window is further divided into 50ms segments, where each segment is flattened into a 1D vector, forming a single column in the resulting 2D matrix. This results in a matrix with 100 columns (corresponding to the 100 segments of 50ms each), while the number of rows depends on the sampling frequency of the EEG signal. The sampling frequency of the EEG in Fig.~\ref{fig_dataprep}(a) is 500Hz. Each resulting 2D image, therefore, represents precisely 5 seconds of EEG data, maintaining both spatial and temporal continuity. To ensure compatibility with standard CNN architectures, the 2D matrix is reshaped into a 224$\times$224$\times$3 RGB image, making it a structured and information-rich input format for feature extraction. Each 50\,ms segment vector is normalized to [0, 1] and mapped to RGB channels using a perceptually uniform colormap (example ‘viridis’ in Matplotlib). 256 discrete color levels map EEG values to 8-bit RGB intensities. Note that depending on data availability, 2-second windows with 20ms segments, with other colormaps may be used to produce the 2D image structure while maintaining compatibility with the framework.

Fig.~\ref{fig_dataprep}(b) shows examples of raw EEG waveforms and the corresponding 2D images for various classes in the BCI~IV and TUH datasets. These image representations reveal class-specific distinctions more clearly than the raw waveforms alone. By introducing this image-based representation as input, we can use our \emph{SSL-SE-EEG} framework to extract meaningful features more effectively.

\subsubsection{\textbf{Modified SSL implementation with SE-Net}}

Our {\emph{SSL-SE-EEG} framework is inspired by Simple Contrastive Learning (SimCLR), a widely adopted self-supervised learning method \cite{chen2020simple} and comprises two main components: a pretraining block and a validation block, as illustrated in Fig.~\ref{fig_workflow}.

\paragraph{{\textbf{Stage 1: Pretraining}}}
In the pretraining stage, we first augment each input image to different views of input data. We transform our EEG data to generate multiple views of the same signal like rotation, blur, etc. Next, we use a modified ResNet as the base encoder. Unlike conventional ResNet architectures, we integrate Squeeze-and-Excitation (SE) blocks after each convolutional layer. SE blocks enhance feature learning by adaptively recalibrating channel-wise representations. This mechanism selectively emphasizes informative channels while suppressing less relevant ones, a key advantage over standard convolutions that treat all input channels uniformly.
\begin{figure}[htpb]
    \centering
    ~\vspace{-6mm}
    \includegraphics[width=1\linewidth]{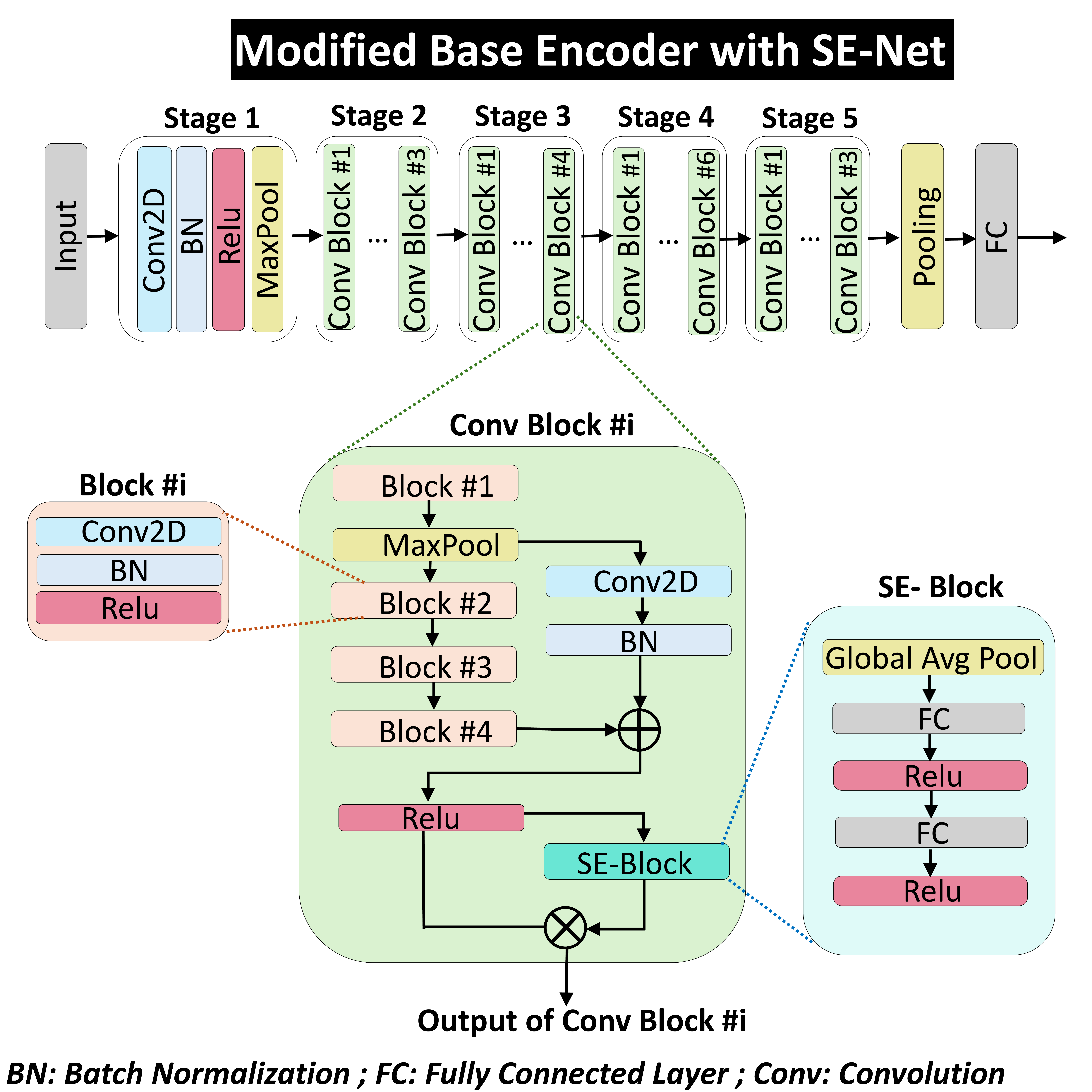}
    ~\vspace{-2mm}
    \caption{Modified Base Encoder with SE-Net. SE-blocks enhance EEG feature sensitivity via global pooling and fully connected layers.}
    ~\vspace{-4mm}
    \label{f_se}
\end{figure}

Fig.~\ref{f_se} illustrates the SE block architecture, where each SE block operates in three stages as follows: 

\begin{itemize} \item \textbf{Squeeze}: A global pooling operation aggregates information across the spatial dimensions, capturing the overall feature distribution.
\item \textbf{Excitation}:  Two fully connected layers learn the relative importance of each channel, highlighting discriminative features.
\item \textbf{Recalibration}: The learned channel weights rescale the original feature maps, suppressing noise while enhancing relevant EEG signals.
 \end{itemize}

By integrating SE blocks with SSL, the framework automatically refines EEG representations, improving robustness against sensor dropouts and signal artifacts. 

Following the encoder, a projection head (implemented as a multi-layer perceptron) maps features into a lower-dimensional space. This step facilitates contrastive learning by ensuring that different transformations of the same EEG segment are mapped closer together, while representations of distinct EEG segments are pushed apart.

The framework is trained using NT-Xent loss (Normalized Temperature-Scaled Cross Entropy), a contrastive loss function shown in Eq~\eqref{eq_loss}~\cite{chen2020simple}. This objective maximizes similarity between augmented versions of the same EEG signal while ensuring separation from other samples.
\begin{equation}
\label{eq_loss}
\mathbb{L}_{i,j} = -\log \frac{\exp(\text{sim}(\mathbf{z}_i, \mathbf{z}_j)/\tau)}
{\sum\limits_{k=1}^{2N} \mathbf{1}_{[k \neq i]} \exp(\text{sim}(\mathbf{z}_i, \mathbf{z}_k)/\tau)}\\
\end{equation}
where $\mathbf{z}_i$ \& $\mathbf{z}_j$ denote latent embeddings of EEG samples, ${sim}$ is cosine similarity,$\tau$ is the temperature parameter and  $N$ is the batch size


Through contrastive learning and channel-wise recalibration, \emph{SSL-SE-EEG} enables noise-tolerant feature learning. This approach significantly enhances model robustness and improves classification performance on downstream EEG tasks. By leveraging SSL pretrain, our framework effectively learns discriminative EEG representations, even in scenarios with limited labeled data.

\paragraph{{\textbf{Stage 2: Validation}}}

After pretraining, is the validation phase, where the encoder's weights remain frozen. The learned feature representations are then used for classification on a labeled EEG dataset. A classification head is added on top of the frozen encoder and trained using categorical cross-entropy loss to assess the model's performance.




\section{Experimental Setup}
\label{sec_setup}


Using multiple public EEG datasets, we evaluate \emph{SSL-SE-EEG} on NVIDIA L40 GPUs. Data preprocessing was performed using Python libraries such as \texttt{Pandas}, \texttt{NumPy}, and \texttt{OpenCV}, while model training and evaluation were conducted with \texttt{TensorFlow}. The model performance was evaluated using standard evaluation metrics such as accuracy and F1 score and computed with \texttt{Scikit-learn}. Accuracy measures the proportion of correctly predicted samples, providing a general indicator of model performance. However, in imbalanced datasets, accuracy alone may be misleading. The F1 score, a harmonic mean of precision and recall, offers a more reliable performance measure across all classes, as shown in (Eq. ~\eqref{eq_acc}).

\begin{subequations} \label{eq_acc}
~\vspace{-3mm}
\begingroup
\setlength{\arraycolsep}{1.5pt} 
\begin{equation}
\textstyle \text{Accuracy} = \frac{TP + TN}{TP + FP + TN + FN}
\end{equation}
\endgroup
~\vspace{-5mm}
\begingroup
\setlength{\arraycolsep}{1.5pt} 
\begin{equation}
\textstyle F1 = \frac{2TP}{2TP + FP + FN}
\end{equation}
\endgroup
\end{subequations}




We use the following datasets:

\begin{enumerate}
\item \textbf{MindBigData(MBD).} This dataset comprises 2-second EEG recordings collected from commercial headsets (NeuroSky MindWave, Emotiv EPOC, Interaxon Muse, and Emotiv Insight). The participants viewed and mentally processed the digits (0-9), forming a 10-class classification task where each class corresponds to a digit ~\cite{MindBigD81:online}.

\item \textbf{TUH EEG Abnormal Corpus (TUH-AB).} This is one of the most extensive publicly available clinical EEG datasets. It contains manually labeled recordings, leading to a two-class classification task in which the classes represent normal and abnormal EEG patterns. Each recording lasts approximately 20 minutes and is captured from 21 EEG channels~\cite{obeid2016temple}.

\item \textbf{SEED-IV.} This dataset comprises EEG recordings of 15 participants watching videos designed to evoke emotions. The recordings are categorized into four distinct emotions: happy, sad, neutral, and fear~\cite{SEEDData42:online}.

\item \textbf{BCI-IV.} This dataset includes EEG recordings from healthy participants engaged in motor imagery tasks. Data was continuously recorded using 59 channels of an Ag/AgCl electrode cap, as subjects envisioned moving their left hand, right hand, or foot, forming a 3-class classification task~\cite{blankertz2007non}.

\end{enumerate}

\section{Evaluation and Results}

In this section, we evaluate our proposed framework \emph{SSL-SE-EEG} to address the following:

\begin{enumerate}
    \item \textbf{Performance of SSL-SE-EEG:} How accurately does the framework classify EEG signals on public datasets? 
    \item \textbf{Accuracy benefit of SE-Net:} How does incorporating SE-Net enhance feature representation and improve model performance?
    \item \textbf{Impact on Power Consumption with SE-Net:} Does including SE-Net significantly impact power consumption, and how does this affect the feasibility of ultra-low-power wearable EEG systems?
\end{enumerate}
This section is divided into three parts, each addressing one of the above questions and offering an analysis of the results and insights gained.

\begin{table*}[t]
\centering
\caption{Comparison with State-of-the-Art Methods on MindBigData and TUH-AB}
\label{table:comparison_sota}
\begin{tabular}{lcccc}
\toprule
\textbf{Paper} & \textbf{Preprocessing} & \textbf{Dataset} & \textbf{Architecture} & \textbf{Accuracy (\%)} \\
\midrule
\textbf{SSL-SE-EEG (Ours)}            & 2D Image Representation & MindBigData & SSL-SE-EEG           & 91  \\
\cite{kumari2021convolutional}         & Spectrogram             & MindBigData & CNN               & 91  \\
\cite{mahapatra2023eeg}                & Raw EEG (Time Series)                 & MindBigData & DWT + BiLSTM      & 71  \\
\cite{falciglia2024learning}           & Spectrogram     & MindBigData & CNN               & 86  \\
\textbf{SSL-SE-EEG (Ours)}                 & 2D Image Representation of EEG & TUH-AB & SSL-SE-EEG      & 85.18    \\
\cite{roy2018deep}                     & Raw EEG (Time Series)                            & TUH-AB & 1D-CNN-RNN  & 82.27 \\
\cite{kamsvaag2023exploring}           & Raw EEG (Time Series)                            & TUH-AB & SSL         & 84.26 \\
\cite{soccol2022attention}            & Raw EEG (Time Series)                            & TUH-AB & LSTM + Attention & 74 \\
\bottomrule
\end{tabular} ~\vspace{-4mm}
\end{table*}

\subsection{Prediction Accuracy of \emph{SSL-SE-EEG}}
We evaluate the performance of our proposed framework, \emph{SSL-SE-EEG}, using a two-phase approach: pretraining followed by downstream fine-tuning, as described in Section~\ref{sec_method}. We conduct experiments on two distinct EEG datasets: MBD \cite{MindBigD81:online}, which is imbalanced with 11 classes, and TUH-AB \cite{obeid2016temple}, which is balanced with three classes, introduced in Section~\ref{sec_setup}.

\textbf{Experiment 1: Pretraining on MBD.} We pretrained \emph{SSL-SE-EEG} on 50{,}000 images from the imbalanced MBD dataset for 50 epochs, then fine-tuned it on two tasks: (1) an unseen 2{,}500-image subset from MBD, and (2) 2{,}500 images from TUH-AB. The model achieved 91.12\% and 89.24\% accuracy, respectively, demonstrating robust feature learning from imbalanced data and across datasets.

\textbf{Experiment 2: Pretraining on TUH-AB.} Next, we pretrained on 50{,}000 images from the balanced TUH-AB dataset, then fine-tuned the model on an unseen subset of 2{,}500 TUH-AB images and 2{,}500 MBD images. This approach yielded slightly lower accuracy: 86.43\% on TUH-AB and 85.18\% on MBD, suggesting that despite good performance, further adaptation is needed to more effectively address class imbalance.

\textbf{Insights.} These experiments reveal two key insights. First, \emph{SSL-SE-EEG} generalizes well across EEG datasets with varying class distributions, confirming the robustness and transferability of its learned representations. Second, even when fine-tuned on limited data, the framework achieves performance comparable to the state-of-the-art methods, as shown in Table~\ref{table:comparison_sota}. These findings underscore the efficiency and broad applicability of \emph{SSL-SE-EEG} in EEG analysis tasks.

\begin{table*}[ht]
\centering
\caption{Comparison of classification performance with and without SE using both Supervised and SSL methods.}
\label{table:SE_results}
\renewcommand{\arraystretch}{1.1}
\resizebox{\textwidth}{!}{%
\begin{tabular}{lcccccccccc}
\toprule
\textbf{Dataset} & \textbf{Type} & \textbf{\# Classes} 
& \multicolumn{2}{c}{\textbf{Without SE - Supervised}} 
& \multicolumn{2}{c}{\textbf{With SE - Supervised}}  
& \multicolumn{2}{c}{\textbf{Without SE - SSL}} 
& \multicolumn{2}{c}{\textbf{With SE - SSL}} \\

\cmidrule(lr){4-5} \cmidrule(lr){6-7} \cmidrule(lr){8-9} \cmidrule(lr){10-11}
& & & \textbf{Acc (\%)} & \textbf{F1} 
& \textbf{Acc (\%)} & \textbf{F1} 
& \textbf{Acc (\%)} & \textbf{F1} 
& \textbf{Acc (\%)} & \textbf{F1} \\
\midrule
\textbf{MindBigData} & Imbalanced & 11 & 57 & 0.45 & 72 & 0.69 & 71 & 0.64 & 84 & 0.92 \\
\textbf{BCI IV}      & Balanced   & 3  & 55 & 0.53 & 62 & 0.61 & 64 & 0.64 & 72 & 0.76 \\
\textbf{SEED IV}     & Imbalanced & 4  & 68 & 0.65 & 78 & 0.70 & 73 & 0.72 & 83 & 0.78 \\
\textbf{TUH-AB}      & Balanced   & 2  & 62 & 0.60 & 70 & 0.69 & 68 & 0.67 & 75 & 0.75 \\
\bottomrule
\end{tabular}%
 } ~\vspace{-4mm}

\end{table*}

\begin{figure} [t]
    \centering
    \includegraphics[width=1\linewidth]{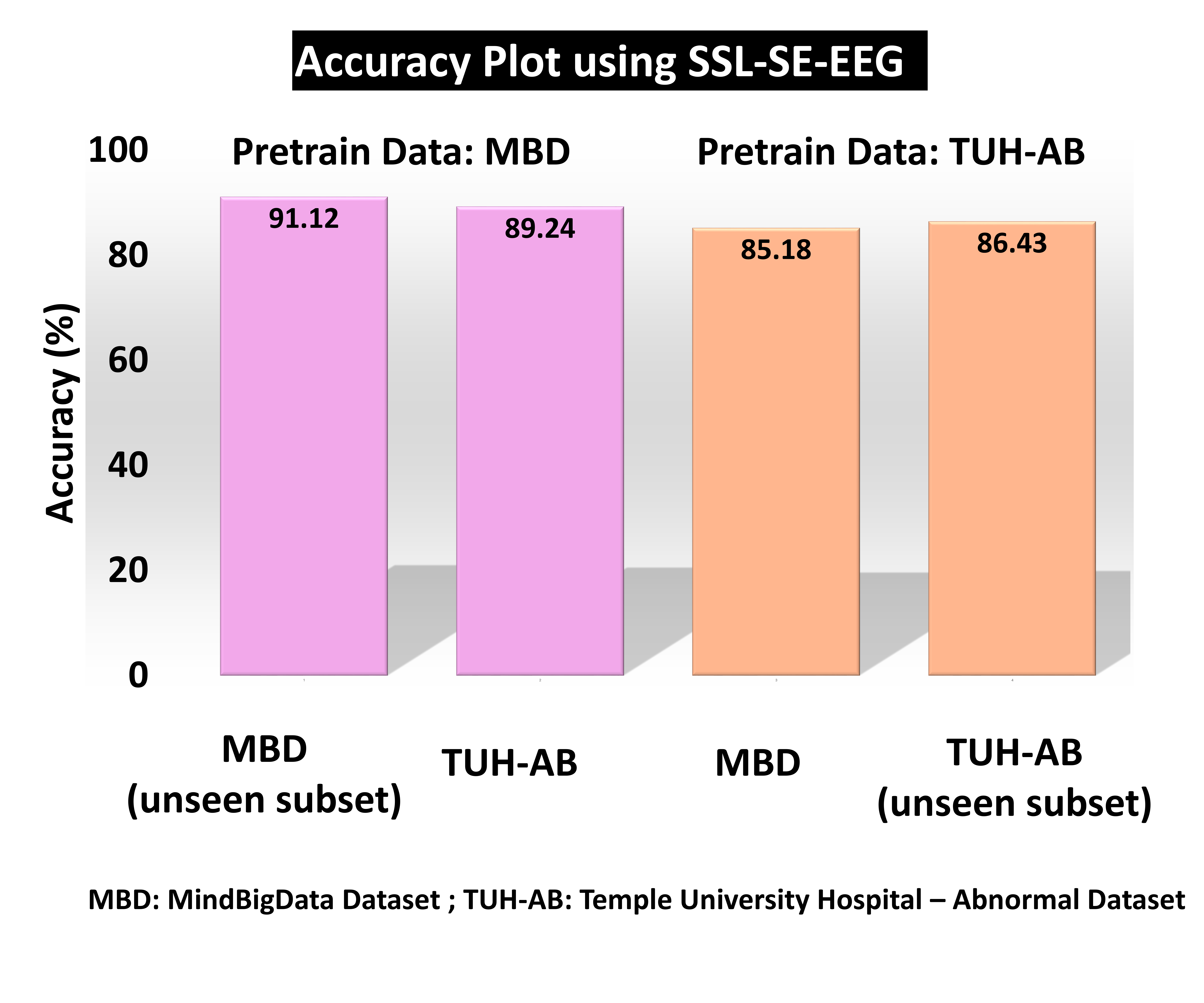}
    ~\vspace{-14mm}
    \caption{Generalization performance of \emph{SSL-SE-EEG} across datasets. The model is pretrained on MBD or TUH-AB and tested on unseen subsets, showing strong cross-dataset transfer with minimal accuracy drop.}
     ~\vspace{-4mm}
    \label{fig_accssl}
\end{figure} 
\subsection{Accuracy benefit of SE-Net}

To assess the benefit of incorporating the SE module within \emph{SSL-SE-EEG}, we performed experiments on four public EEG datasets. MindBigData~\cite{MindBigD81:online}, BCI IV~\cite{blankertz2007non}, SEED IV~\cite{SEEDData42:online}, and TUH-AB~\cite{obeid2016temple}. We used a 80\%/20\% split for pre-training and downstream validation for each data set, respectively. To ensure a fair comparison, all models were pre-trained for 10 epochs using the 2D image representation described in Section~\ref{sec_method}, and performance was evaluated using accuracy and F1 score (Equations \ref{eq_acc}(a) and (b)).

Table~\ref{table:SE_results} compares models with and without the SE module under both SSL-pretrained and SL conditions. The integration of the SE module consistently boosts performance across all datasets. For example, on the MBD imbalanced data set, the SSL-pretrained model accuracy improved from 71\% to 84\%, and its F1 score increased from 0.64 to 0.92 with SE. Similar improvements were observed in the balanced BCI IV and TUH-AB datasets and the imbalanced SEED IV datasets.

{\textbf{Insights.} 
Performance gains can be attributed to the SE module's ability to dynamically recalibrate channel-wise feature responses, effectively allowing the network to focus on the most informative and discriminative features. On imbalanced datasets like MindBigData and SEED IV, where minority classes are often underrepresented, this channel reweighting leads to more balanced attention across classes, thereby improving both accuracy and F1 score. In contrast, balanced datasets such as BCI IV and TUH-AB benefit from the SE module through enhanced feature extraction and noise suppression, which results in more robust classification even under varying conditions. These improvements are evident regardless of whether SSL pre-training is used, underscoring the broad applicability of the SE module and its crucial role in enhancing EEG signal analysis.



\subsection{Impact on Power Consumption by incorporating SE-Net}


\begin{figure} [t]
    \centering
    \includegraphics[width=1\linewidth]{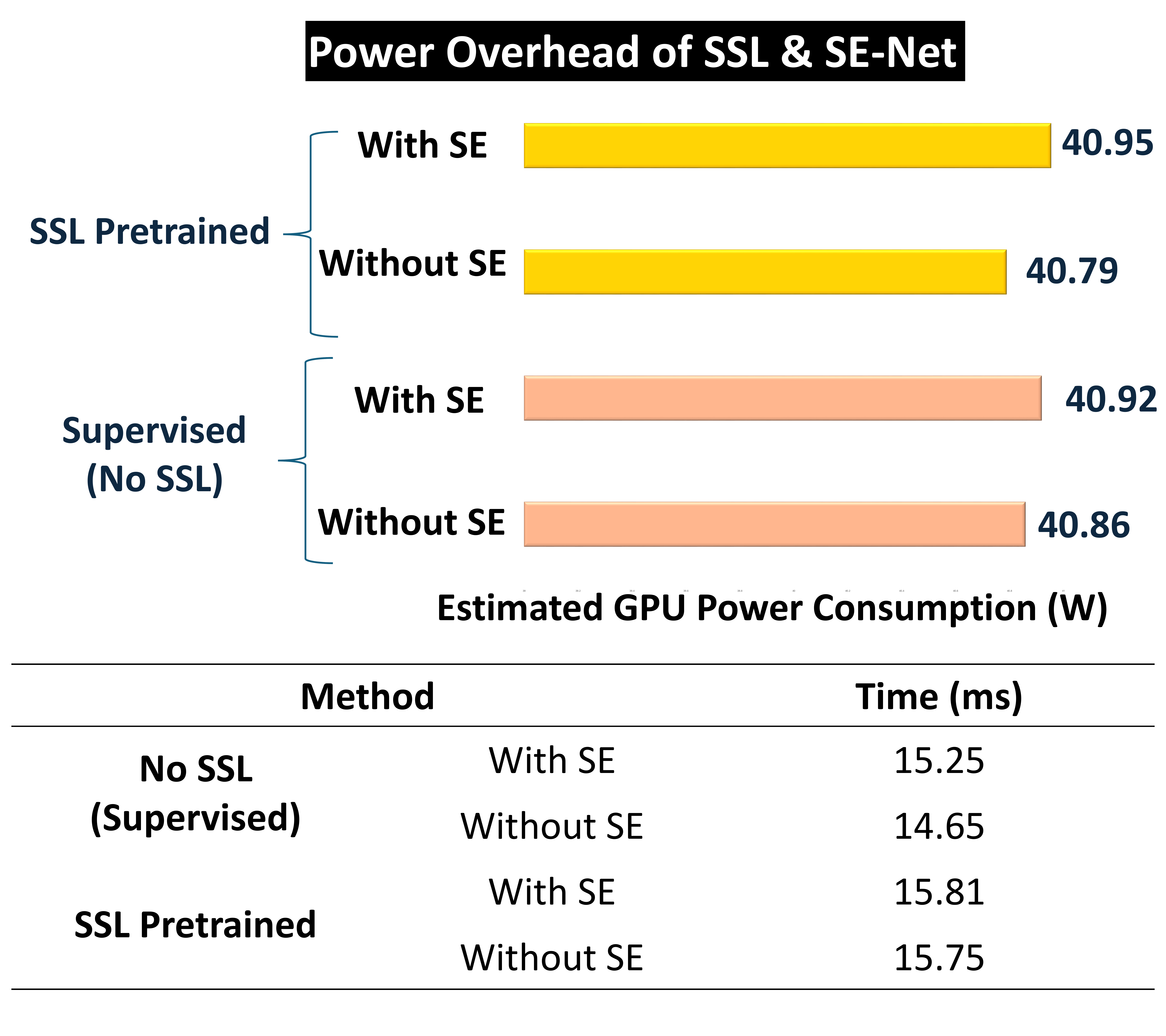}
    ~\vspace{-9mm}
    \caption{GPU power and inference time comparison for \emph{SSL-SE-EEG}. SE-Net slightly increases power ($\leq$0.4\%) and inference time ($\leq$1ms) while enhancing feature extraction.}
    ~\vspace{-6mm}
    \label{fig_power}
\end{figure}

Next, we examine how adding the SE module affects GPU power consumption and energy use. We evaluated one datapoint from the TUH-AB dataset on one L40 GPU and estimated power using NVIDIA’s NVML library. Fig. \ref{fig_power} illustrates the average GPU power consumption and total energy use under two learning paradigms: SSL pre-trained and SL, comparing setups with and without SE-Net.

In the SL setup, adding SE-Nets leads to a 0.15\% increase in average GPU power. When SSL pretraining is used, adding SE raises the average GPU power by 0.4\%. This shows that the addition of SE-Nets introduces a slight increase in power, attributed to the lightweight nature of global average pooling and simple channel-wise recalibration. The difference in power consumption between SL and SSL is mainly due to their underlying computational methodology rather than the SE module itself. In SL, each sample is processed once through a single ResNet-based CNN, whereas SSL processes multiple augmented views of the same image (see Fig~\ref{fig_workflow}), increasing the number of forward passes and computations per sample. As a result, the same SE module causes a relatively larger increase in power consumption in SSL due to the higher number of computation per input.

{\textbf{Insights.} From a broader perspective, the minor increase in power consumption and runtime compared to non-SE implementations suggests that SE-Nets introduce minimal power overhead while enhancing efficiency. In fact, SE-Nets improve feature extraction and yield more robust representations, leading to more accurate predictions and better generalization. When deployed on ML-optimized hardware, such as~\cite{moons20160}, which operates at an efficiency of 0.3–2.6 TOPS/W, our proposed \emph{SSL-SE-EEG} framework would consume as little as 2.96 mW in the best-case scenario and up to 25.67 mW in the worst case, which is well within the power constraints of mobile and wearable devices. Furthermore, quantization and dedicated hardware can further optimize power efficiency while maintaining performance \cite{chowdhury2024leveraging}. Future work will explore these hardware-aware optimizations to enhance the practicality of \emph{SSL-SE-EEG} in real-world low-power applications. 

\section{Conclusion}
This work introduces \emph{SSL-SE-EEG}, a novel framework that leverages SSL to reduce dependency on labeled data and uses SE-Nets to enhance feature selection and noise suppression. Our approach achieves 91\% accuracy in MindBigData and 85.18\% in TUH-AB. We also perform experiments to demonstrate that SE-Net integration improves classification by up to 15\% public data sets while keeping power consumption low ($\leq$0.4\% compared to no SE-Net). These results demonstrate that \emph{SSL-SE-EEG} enables robust, scalable EEG-based interfaces, making it well-suited for real-time cognitive state monitoring, BCIs, and neurorehabilitation applications. Future work will focus on further optimizing real-time deployment for low-power wearable EEG systems. 



\bibliographystyle{ieeetr}

\end{document}